\documentclass[11pt]{article}
\usepackage[margin=1in]{geometry}
\usepackage{caption}
\usepackage{graphicx} 
\usepackage{array}
\usepackage{booktabs}
\usepackage{tabularx}
\usepackage{url}
\usepackage{hyperref}
\usepackage{amsmath}

\title{ChemGraph-XANES: An Agentic Framework for XANES Simulation and Curation}

\author{
Vitor F. Grizzi$^1$, Thang Duc Pham$^2$, Luke N. Pretzie$^1$, Jiayi Xu$^1$, Murat Keceli$^2$, and Cong Liu$^1$\thanks{Corresponding author. Email: congliu@anl.gov} \\[0.5em]
$^1$Chemical Sciences and Engineering Division \\
Argonne National Laboratory, Lemont, IL 60439, USA \\[0.3em]
$^2$Computational Science Division \\
Argonne National Laboratory, Lemont, IL 60439, USA
}

\begin{document}

\maketitle
\begin{abstract}
Computational X-ray absorption near-edge structure (XANES) is widely used to interpret local coordination environments, oxidation states, and electronic structure, but large computational campaigns are often limited by workflow complexity. We present ChemGraph-XANES, a large language model (LLM)-based agentic framework that combines documentation-grounded parameter retrieval via retrieval-augmented generation (RAG), schema-constrained tool execution, deterministic FDMNES input generation, Parsl-backed execution, and provenance-aware data curation. Scripted and natural-language interfaces share a common scientific backend for structure handling, parameterization, execution, spectral extraction, and optional post-processing. We evaluate three workflow modes: documentation-grounded parameter propagation, structure-file-based execution, and composition-based execution from a chemistry-level request. Repeated trials yielded end-to-end completion in 10/10 composition-based runs, 10/10 structure-file-based runs, and 9/10 documentation-grounded RAG runs. In every RAG run, the energy-grid specification retrieved from the FDMNES manual was correctly propagated, with the single end-to-end failure occurring downstream during multi-structure handling. In a separate task-parallel demonstration, the framework retrieved 21 TiO$_2$ structures from the Materials Project and submitted one FDMNES calculation per structure. All calculations completed successfully, with Parsl distributing the independent tasks across the user-configured worker pool. Together, these results show that ChemGraph-XANES provides a constrained and reproducible orchestration layer for computational spectroscopy, supporting consistent execution of representative tasks, documentation-linked parameter selection, and task-parallel generation of structure-linked XANES collections.

\end{abstract}

\section{Introduction}
\label{sec:introduction}
Computational X-ray absorption near-edge structure (XANES) has become an important complement to experimental X-ray absorption spectroscopy for interpreting local coordination environments, oxidation states, and electronic structure in chemically complex systems \cite{xu2024theoretical,zhu2021k,piquer2014fe,xu2024recent, xu2024recent2}. In practice, however, the primary obstacle to routine, large-scale XANES calculations often lies not only in the electronic-structure method or scattering formalism, but also in the workflow required to execute each calculation in a reproducible manner. A typical XANES campaign may require retrieving or preparing candidate structures, identifying absorber sites, specifying calculation parameters, generating code-specific inputs, organizing many independent runs, parsing spectral outputs, applying consistent normalization, and preserving the provenance linking each spectrum to its parent structure. When these steps are handled manually or through ad hoc scripts, the workflow becomes difficult to reproduce, reuse, and scale.

This workflow bottleneck becomes especially important in settings that require many calculations rather than a single forward simulation. Such settings are common across computational chemistry tasks, including high-throughput screening of structural databases, comparison of DFT-optimized candidate geometries, sampling of thermally accessible configurations from molecular dynamics or \textit{ab initio} molecular dynamics trajectories, iterative refinement against experiments, materials discovery, and dataset generation for machine-learning (ML) models \cite{aldossary2024silico, friederich2021machine, grizzi2026active, choudhary2020joint, jha2019enhancing, ahmad2018machine, sprueill2023monte, xu2026rational, vitillo2026accelerating}. In computational spectroscopy, these same pressures appear when large collections of spectra must be generated and curated consistently for screening, ensemble analysis, experiment-theory comparison, or ML applications \cite{grizzi2026xane3, smith2017soft}. In such settings, the practical value of computational XANES depends not only on the fidelity of the underlying physics-based method, but also on whether structure preparation, parameter specification, execution, post-processing, and data organization can be carried out consistently across large batches of jobs. Because individual XANES calculations are naturally independent once their inputs are defined, these workloads are also well suited to task-parallel execution on high-performance computing (HPC) systems.

Recent progress in large language models (LLMs) and LLM-based agents has created new opportunities for scientific workflow automation and for making scientific workflows more machine-actionable, accessible, and reusable~\cite{yao2022react, luo2025large, wang2024mixture, yang2025multi, zou2025agente, pyzer2025foundation, shen2026unlocking, choudhary2024atomgpt, leong2025steering}. By exposing established domain software as callable tools, high-level natural-language requests can be translated into structured computational actions, reducing the need for users to manually compose each step of the simulation pipeline \cite{takahara2025accelerated, pham2026chemgraph, wang2025dreams, mcnaughton2024cactus}. In computational spectroscopy, this paradigm enables an agent to select the appropriate operation, populate its parameters through typed interfaces, and orchestrate multi-step workflows spanning structure retrieval, simulation setup, execution, and post-processing. Rather than replacing the underlying spectroscopy engine, the agent functions as an orchestration layer over conventional scientific software, improving accessibility while preserving transparent execution logic \cite{lu2026towards}.

In this work, we develop \textbf{ChemGraph-XANES} as a domain-specific extension of the original ChemGraph framework~\cite{pham2026chemgraph} for computational XANES simulation and curation. Built on ASE, FDMNES, Parsl, and a LangGraph/LangChain-based tool interface \cite{ase-paper, bunuau2009self, joly2001x, babuji2019parsl}, the framework exposes XANES workflow operations as typed Python tools, enabling language-model agents to interpret user requests, map relevant quantities onto structured function arguments, and execute the resulting spectroscopy workflow. The central contribution is a documentation-grounded, schema-constrained, provenance-aware orchestration layer that connects structure acquisition, absorbing-species specification, FDMNES input generation, execution, spectral parsing, normalization helpers, and structure-spectrum linkage within a single Python framework. The workflow can be invoked either directly from scripts or through natural-language requests mapped onto typed tool calls, thereby preserving a common deterministic scientific backend across multiple user-facing interaction modes. We demonstrate documentation-grounded parameter retrieval and propagation, support for both local structure-file inputs and chemistry-level requests, task-specific consistency across ten separate executions of each of three representative workflows, and a 21-structure task-parallel TiO$_2$ calculation on an HPC system. The present work combines proof-of-capability demonstrations with a limited repeated-execution evaluation and one task-parallel collection. However, it is not a comprehensive benchmark of spectral fidelity, retrieval performance, general workflow reliability, or distributed-execution scaling. Taken together, these capabilities establish \textbf{ChemGraph-XANES} as a machine-actionable workflow layer for reproducible and extensible XANES simulation and data curation. 

\section{Methods}
\label{sec:methods}

\subsection{Workflow overview and agentic orchestration}
We developed ChemGraph-XANES as a modular, agent-compatible workflow for automated XANES calculations. The framework organizes natural-language task specification, structure acquisition, FDMNES input preparation, task-parallel execution, spectral post-processing, and provenance tracking as callable Python stages. The underlying scientific operations are implemented as Python functions and exposed to the execution layer as typed tools with schema-defined arguments. In agentic mode, a large language model (LLM) interprets a user query, selects the relevant tool, and maps user-specified quantities, such as the chemical system, absorbing species, cluster radius, electronic configuration settings, or output location, onto structured function arguments. These arguments are then passed to the same underlying Python routines used in direct scripted execution, thereby preserving a single source of scientific logic while enabling the workflow to be driven from natural-language requests. In the current implementation, execution, database expansion, and normalization/plotting remain separate callable stages, so an agent trace should not be interpreted as automatically completing every downstream curation step unless those tools are explicitly invoked.

The current implementation supports both single-agent and multi-agent execution modes. In the single-agent setting, the model iteratively alternates between reasoning and tool invocation until the requested workflow is completed. In the multi-agent setting, a planner first decomposes the user request into subtasks, worker agents execute those subtasks through the same tool abstraction, and an aggregator combines the resulting outputs into a final response. Further details of the underlying ChemGraph framework and single- or multi-agent settings are available in Ref. \cite{pham2026chemgraph}. 

\begin{figure}
    \centering
    \includegraphics[scale=2.5]{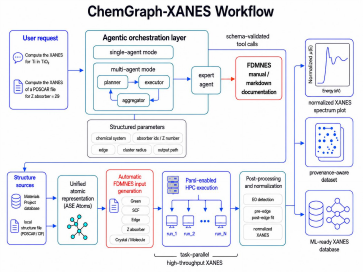}
    \caption{Schematic overview of the ChemGraph-XANES workflow. User requests can be provided either as chemistry-level natural-language queries or as explicit local structure files. These inputs are interpreted by an agentic orchestration layer that supports both single-agent and multi-agent execution, with an optional documentation-grounded expert agent that consults the FDMNES manual to inform parameter selection. The resulting schema-validated tool calls are mapped onto structured arguments, enabling automated structure retrieval, conversion to a unified \texttt{ASE Atoms} representation, FDMNES input generation, and task-parallel execution through \texttt{Parsl} on HPC systems. After successful calculations, separate post-processing helpers can parse spectra, apply normalization when the energy mesh is compatible with the selected fitting windows, and attach spectra back to the corresponding structures for downstream analysis and ML-oriented database generation.}
    \label{fig:general_arch}
\end{figure}

The multi-agent mode can additionally incorporate a retrieval-augmented expert agent that grounds XANES parameter selection in external documentation, such as a text-converted software manual, allowing executor agents to receive documentation-informed guidance without hard-coding all decision rules directly into the executor prompt or the simulation backend. This design enables the same XANES workflow to be invoked from interactive natural-language sessions, notebooks, or conventional Python scripts without changing the underlying spectroscopy routines. At a high level, the pipeline begins with either user-provided structures or structures retrieved from a materials database, converts them into a unified atomistic representation, generates FDMNES input files, executes the calculations, parses the resulting convolution spectra, and stores the spectra back alongside the originating structures for downstream analysis. Figure~\ref{fig:general_arch} provides a schematic overview of the ChemGraph-XANES workflow.

It is useful to distinguish three levels of capability provided by the current implementation. First, at the deterministic workflow level, the framework supports conventional scripted execution of structure handling, FDMNES input generation, execution, parsing, and optional post-processing without requiring any language-model interaction. Second, at the orchestration level, a natural-language request can be translated into schema-validated tool calls that populate the same underlying scientific routines. Third, at the documentation-grounded assistance level, external reference material can be retrieved and used to inform parameter selection before execution. This layered design is intended to separate the benefits of language-based interaction from the underlying scientific logic, which remains implemented in explicit Python functions and constrained tool schemas rather than in free-form model output. Accordingly, ChemGraph-XANES should not be interpreted as granting unrestricted autonomy to a language model over arbitrary FDMNES inputs. Instead, the agent operates within constrained interfaces that specify which parameters can be set, how they are validated, and how they are passed to the backend. This distinction is important for reproducibility and trustworthiness, particularly in scientific settings where undocumented parameter changes can compromise later interpretation or reuse.

\subsection{Structure acquisition and curation}
For database-driven workflows, bulk structures are retrieved programmatically from the Materials Project through its Python API \cite{jain2013commentary, ong2015materials}. The current implementation queries the materials summary endpoint using a user-specified chemical formula together with an energy-above-hull threshold in order to restrict the search to low-energy candidate structures. The retrieval tool returns all structures satisfying the filter and does not itself reduce a formula-level query to a single structure. When a request requires one structure, selection from the returned candidates occurs during orchestration unless the user supplies an explicit structure identifier or selection rule; requests that explicitly ask for all qualifying structures retain the complete returned list. Retrieved structures are converted from the native Pymatgen representation into \texttt{ase.Atoms} objects, which are used as the internal structural representation throughout the workflow. Each structure is annotated with its Materials Project identifier, thereby preserving traceability to the original source and simplifying later aggregation of results.

To support both interoperability and local reuse, structures retrieved from the Materials Project are stored in two complementary formats. Each structure is exported as an individual \texttt{.cif} file, preserving compatibility with standard crystallographic tools, and the full retrieved collection is serialized as a pickled list of \texttt{ase.Atoms} objects in \texttt{atoms\_db.pkl}. In the \texttt{create\_fdmnes\_inputs} helper, this serialized collection is used as the fallback source for FDMNES input generation when an explicit in-memory structure list is not supplied. Other execution paths can also operate directly on supported structure files, directories, or ASE databases.

\subsection{Automatic FDMNES input generation}
Automatic FDMNES input generation in ChemGraph-XANES combines deterministic script-based file construction with an agentic parameter-selection layer. Within the broader ChemGraph architecture, a retrieval-augmented agent can ingest text or PDF-derived documentation, index its contents into a queryable knowledge base, and retrieve grounded guidance when an FDMNES calculation must be parameterized. The knowledge base is constructed locally at runtime from user-supplied documentation rather than stored as a fixed internal resource; users can reconstruct and query it using the shared code and the corresponding source document. In the demonstration trace reported below, the RAG agent graph exposes both documentation-retrieval tools and XANES execution tools, allowing a retrieved \texttt{Range} value to be passed into a schema-validated XANES tool call. This should be interpreted as a configured tool-level coupling demonstration rather than as a fully general expert/executor controller that automatically runs all downstream normalization and data-curation helpers. Agent decisions are validated against structured schemas, which translate high-level reasoning into the standardized parameters required by the backend.

Once the calculation arguments are fixed, FDMNES input files are generated directly from the \texttt{ase.Atoms} objects. For each structure, the workflow writes both the steering file \texttt{fdmfile.txt} and the main input file \texttt{fdmnes\_in.txt}. The absorber can be specified either by atomic species or by site index. If the user provides \texttt{absorber\_idx}, the workflow writes the FDMNES \texttt{Absorber} keyword with that 1-based atom index, enabling a site-specific calculation for the selected atom in the structure ordering. If \texttt{absorber\_idx} is not provided but \texttt{z\_absorber} is specified, the workflow writes \texttt{Z\_absorber}, which selects all atoms with that atomic number as the absorbing species in the FDMNES sense. If neither argument is supplied, the workflow chooses the element with the largest atomic number present in the structure as a deterministic fallback and writes it through \texttt{Z\_absorber}. Edge selection is exposed through the \texttt{edge} argument, which is written to the generated input with the FDMNES \texttt{Edge} keyword and defaults to the K edge when no other edge is specified.

The FDMNES energy mesh can be supplied through the schema-validated \texttt{energy\_range} argument, either explicitly by the user or through documentation-grounded retrieval from the FDMNES manual. More generally, the exposed FDMNES controls, including \texttt{energy\_range}, \texttt{radius}, \texttt{edge}, \texttt{magnetism}, \texttt{Green}, \texttt{Density\_all}, \texttt{Quadrupole}, \texttt{Spherical}, and \texttt{SCF}, have implementation-level defaults but can be overridden through schema-validated arguments when specified by the user or selected with documentation-grounded expert-agent guidance.

The current ChemGraph-XANES implementation defaults to a cluster radius of 6.0~$\mathrm{\AA}$ and a variable energy mesh of \texttt{[-55, 1, -10, 0.01, 5, 0.1, 150]}~eV. Calculations default to the K edge, with \texttt{Green}, \texttt{Density\_all}, \texttt{Quadrupole}, \texttt{Spherical}, and \texttt{SCF} enabled and \texttt{Magnetism} disabled. These values are ChemGraph-XANES implementation defaults rather than defaults supplied by FDMNES. Unless otherwise specified, they were used for the single-structure calculations reported here. Spectral convolution was requested automatically without overriding the broadening parameters supplied internally by the installed FDMNES version. The documentation-grounded RAG calculations used the manual-derived FDMNES \texttt{Range} setting, corresponding to an energy mesh of \texttt{[-5.0, 0.5, 60.0]}~eV, instead of the ChemGraph-XANES default mesh. The task-parallel TiO$_2$ batch used a 5.0~$\mathrm{\AA}$ cluster radius and an energy mesh of \texttt{[-20.0, 1.0, -10.0, 0.05, 10.0, 0.1, 50.0]}~eV. For this batch, \texttt{Green}, \texttt{Quadrupole}, and \texttt{SCF} were enabled, whereas \texttt{Density\_all}, \texttt{Spherical}, and \texttt{Magnetism} were disabled.

The workflow distinguishes automatically between periodic and non-periodic structures. If all periodic boundary condition flags are enabled, the structure is written in \texttt{Crystal} mode using the cell parameters and fractional atomic coordinates. Otherwise, the structure is written in \texttt{Molecule} mode using Cartesian coordinates and a cubic bounding cell constructed from the atomic positions. This branching enables a common front-end representation to be used for both crystalline and molecular systems while still emitting valid FDMNES inputs for each case.

More broadly, the parameter-exposure strategy in ChemGraph-XANES is deliberately conservative. Only a defined subset of FDMNES controls is surfaced through schema-validated interfaces, with implementation-level defaults used when explicit user input or documentation-grounded guidance is not provided. This restriction is intentional: it preserves flexibility for common XANES setup choices while reducing the risk that unconstrained language-model output could introduce unsupported or poorly justified simulation settings. In this sense, the framework uses the agent as a controlled orchestration layer over deterministic input-generation logic, rather than as an unrestricted author of full simulation files.

\subsection{Task-parallel execution, Parsl compatibility, and provenance tracking}
To support batch studies, the pipeline includes a batch input generator that creates a dedicated run directory for each structure under a common \texttt{fdmnes\_batch\_runs} root. Each directory contains the corresponding FDMNES input files together with a serialized copy of the associated \texttt{ase.Atoms} object. The filename of the serialized structure includes the absorbing-species atomic number, the Materials Project identifier when available, and the chemical formula. This naming convention provides a lightweight traceability layer that links each calculation directory to its structural origin.

Single-run execution is handled by a core routine that reads an input structure through ASE, creates the output directory, writes the FDMNES inputs, and launches the external FDMNES executable through a subprocess call. The path to the executable is provided through the \texttt{FDMNES\_EXE} environment variable. Standard output and standard error streams are redirected to \texttt{fdmnes\_stdout.txt} and \texttt{fdmnes\_stderr.txt}, respectively, which facilitates debugging and post hoc inspection of failed jobs. After execution, the workflow searches the run directory for FDMNES convolution output files matching \texttt{*conv.txt}. All matching spectra are parsed and stored as NumPy arrays. A scientifically successful run requires at least one convolution output file; ensemble-level summaries should therefore check this file count explicitly rather than treating completion of a Parsl task alone as evidence of a completed XANES calculation.

Because each structure gives rise to an independent FDMNES calculation once the input parameters are specified, the workflow is naturally task-parallel across structures. This property makes it compatible with distributed execution through \texttt{Parsl}, where independent structure-level jobs can be dispatched on HPC systems while retaining a uniform interface for inputs, logs, and outputs. Because the focus of the present work is workflow construction and representative validation, we evaluate execution and curation under one task-parallel configuration rather than undertaking a dedicated performance-scaling study. Quantitative throughput, speedup, scaling efficiency, and operational failure rates across broader deployments remain targets for future work.

To assemble the results into a dataset suitable for downstream analysis, a separate helper reloads the serialized structures, attaches the extracted spectral data to the corresponding \texttt{ase.Atoms} objects under the \texttt{atoms.info["FDMNES-xanes"]} field, and writes the expanded collection to \texttt{atoms\_db\_expanded.pkl}. This optional post-processing step preserves the association between structural inputs and spectral outputs in a form that can be reused for subsequent analysis, data curation, and ML-oriented dataset construction.

\subsection{Evaluation protocols}
To evaluate task-specific repeated-execution consistency, we performed ten separate executions of each of three representative workflows using Gemini 2.5 Flash with a temperature of 0 and a maximum output of 6000 tokens. The evaluated tasks were (i) composition-based execution using the original prompt \textit{``Compute the XANES for Ti in TiO$_2$.''}, (ii) structure-file-based execution from the supplied POSCAR with $Z_{\mathrm{absorber}}=29$, and (iii) a documentation-grounded RAG workflow that retrieved the default FDMNES \texttt{Range} values from the manual and used them for TiO$_2$ and Fe$_2$O$_3$ calculations. The RAG executions used the document-indexing, embedding, and retrieval configuration described in the documentation-grounded parameter-selection subsection. A trial was counted as an end-to-end success only when every calculation requested by the prompt produced at least one parseable convolution spectrum and manual inspection confirmed that the generated FDMNES inputs contained the requested parameters. Retrieval and propagation of the manual-derived \texttt{Range} setting were also assessed separately from end-to-end completion. This protocol measures observed consistency for the three fixed tasks under the reported configuration rather than general workflow reliability.

The task-parallel evaluation used Gemini 2.5 Flash in the multi-agent XANES configuration. The prompt template is reproduced below; at execution time, \texttt{\{RUNS\_DIR\}} was replaced by the absolute output-directory path.

\begin{verbatim}
Fetch all non-deprecated TiO2 structures from Materials Project with
energy_above_hull <= 0.10 eV/atom. Pass exactly the returned
structure_files list to one run_xanes_ensemble call; do not scan
the output directory or call run_xanes_single.

Run Ti K-edge calculations with:
output_dir="{RUNS_DIR}", z_absorber=22, edge="K", radius=5.0,
energy_range=[-20.0, 1.0, -10.0, 0.05, 10.0, 0.1, 50.0],
green=true, density_all=false, quadrupole=true, spherical=false,
scf=true, magnetism=false, and skip_completed=false.

Do not plot or modify the spectra. Report the retrieved, successful,
and failed structure counts and the absolute path to
xanes_results.jsonl.
\end{verbatim}

The exact structure-file list returned by the Materials Project retrieval tool was supplied to one \texttt{run\_xanes\_ensemble} call. For this experiment, Parsl was configured with a maximum of ten one-node worker blocks, with one worker assigned to each 128-core compute node and one FDMNES calculation executed per worker. This configuration permitted up to ten structure-level calculations to run concurrently.

\subsection{Spectral normalization}
Raw FDMNES convolution outputs are further processed through a normalization routine designed to place spectra on a common scale. Let $E$ denote the energy grid and $\mu(E)$ the raw absorption signal read from a convolution file. The workflow first assigns the edge reference energy as $E_0=0$ when the computed energy grid spans zero, consistent with the FDMNES convention that energies are reported relative to the edge. 

Two linear baselines, $\mu_{\mathrm{pre}}(E)$ and $\mu_{\mathrm{post}}(E)$, are fitted to the pre-edge and post-edge regions, respectively. With the default settings, the pre-edge region contains points satisfying $E \leq E_0-20$~eV, whereas the post-edge region contains points satisfying $E \geq E_0+50$~eV. The pre-edge baseline is subtracted from the spectrum, and the result is normalized by the difference between the two fitted baselines at $E_0$:

\[
\mu_{\mathrm{norm}}(E)=\frac{\mu(E)-\mu_{\mathrm{pre}}(E)}{\mu_{\mathrm{post}}(E_0)-\mu_{\mathrm{pre}}(E_0)}.
\]

Both fitting regions must contain sufficient points for a linear fit. If the supplied energy mesh does not cover either region or a fit fails, the workflow instead applies maximum-intensity normalization:

\[
\mu_{\mathrm{norm}}(E)=\frac{\mu(E)}{\max_E\mu(E)}.
\]

When the linear fits are valid, the baseline-subtracted edge-step procedure yields a normalized absorption curve that is less sensitive to baseline offsets and differences in absolute intensity scale. The maximum-intensity fallback places spectra on a common intensity scale but does not perform pre-edge subtraction or edge-step normalization. Because normalization is not scientifically neutral, ChemGraph-XANES preserves the raw convolution output alongside the normalized spectrum so that later analyses can revisit the fitting windows, edge reference, or normalization convention. Standardized normalization is useful for comparative studies and ML-oriented datasets, but it should be treated as a validated post-processing step rather than as an automatic consequence of every FDMNES run. For convenience, a plotting routine iterates over completed run directories, applies the normalization procedure, and writes an \texttt{xanes\_plot.png} file for each successfully parsed calculation. 

\section{Results}   
\subsection{Documentation-grounded parameter selection and propagation into execution}
A key feature of ChemGraph-XANES is a documentation-grounded expert agent that consults external reference material before downstream tool invocation. Rather than relying exclusively on latent model knowledge, the expert agent retrieves evidence from a local knowledge base constructed from the FDMNES manual and uses that evidence to guide parameter selection. This functionality is especially important for XANES workflows because input parameters often depend on code-specific keywords, defaults, and conventions that are easy to misstate in free-form language.

To illustrate this behavior, we evaluated the expert agent on three representative parameterization queries concerning default absorbing-species selection, substitutional doping, and default energy-range settings. The qualitative retrieval examples in Table~\ref{tab:expert-agent-demo} were generated with GPT-4o, while the executable provenance trace in Table~\ref{tab:range-provenance} was generated with Gemini 2.5 Flash. The use of two providers was intended to demonstrate model portability rather than comparative performance. Document retrieval used the Hugging Face sentence-transformers/all-MiniLM-L6-v2 embedding model on CPU and a FAISS index, with a chunk size of 1,000 characters and an overlap of 200 characters.

In all three cases, the relevant FDMNES guidance was obtained through retrieval rather than supplied directly in the prompt. For readability, Table~\ref{tab:expert-agent-demo} presents condensed traces containing the user query, the explicit retrieval call, the salient evidence extracted from the manual, and a shortened excerpt from the final agent response. The raw interaction logs for the qualitative examples are available in the accompanying repository notebook, and the raw Gemini trace for the range-propagation example is provided separately.\footnote{\url{https://github.com/vitorgrizzi/ChemGraph_xanes/blob/main/notebooks/Demo_rag_agent_Argo.ipynb}; \url{https://github.com/vitorgrizzi/ChemGraph_xanes/blob/main/notebooks/rag_range_fill.txt}}

Across all three cases, the agent first invoked the \texttt{query\_knowledge\_base} tool with a targeted search phrase and then produced a response consistent with the retrieved manual content. The first example is especially instructive because it highlights both the usefulness and the limitations of agent-mediated summarization: the final response captures the practical default behavior and correctly points the user to the \texttt{Z\_absorber} and \texttt{Absorber} keywords, while the retrieved manual evidence provides the more precise FDMNES convention that the calculated spectra correspond to atoms sharing the atomic number of the first species listed under \texttt{Crystal} or \texttt{Molecule}. The second and third examples show that the same retrieval mechanism can recover both procedural instructions, such as the use of the \texttt{Doping} keyword, and default numerical settings, such as the standard energy range and step size. Although these examples are qualitative demonstrations rather than a formal retrieval benchmark, they show that the expert-agent layer can consult external documentation before answering parameterization questions, improving transparency and reducing reliance on unsupported free-form reasoning.

\begin{table*}[!t]
\centering
\caption{Representative documentation-grounded expert-agent retrieval examples. For readability, the retrieved evidence and final answer fields are condensed from the raw execution traces. Full interaction is available at \url{https://github.com/vitorgrizzi/ChemGraph_xanes/blob/main/notebooks/Demo_rag_agent_Argo.ipynb}.}
\label{tab:expert-agent-demo}
\renewcommand{\arraystretch}{1.22}
\setlength{\tabcolsep}{6pt}
\small
\begin{tabular}{>{\raggedright\arraybackslash}p{0.22\textwidth} >{\raggedright\arraybackslash}p{0.72\textwidth}}
\toprule

\multicolumn{2}{l}{\textbf{Example 1}} \\[2pt]
\textbf{User query} & How does the program determine the absorbing species by default? \\[4pt]
\textbf{Tool call} & \texttt{query\_knowledge\_base(query="default absorbing species determination", file\_path=FDMNES\_manual.txt)} \\[4pt]
\textbf{Retrieved evidence} & By default, the calculated spectra correspond to the sum of the scattering or absorption produced by all atoms with the same atomic number as the first one listed under \texttt{Crystal} or \texttt{Molecule}. Explicit overrides are available through \texttt{Z\_absorber} or \texttt{Absorber}. \\[4pt]
\textbf{Agent answer (excerpt)} & ``By default, the calculated spectra correspond to atoms with the same atomic number as the first species listed under \texttt{Crystal} or \texttt{Molecule}. If a different absorbing species is intended, the user can explicitly define the absorbing-species atomic number using \texttt{Z\_absorber}.'' \\[6pt]

\midrule

\multicolumn{2}{l}{\textbf{Example 2}} \\[2pt]
\textbf{User query} & How can a user model substitutional doping in a crystal structure? \\[4pt]
\textbf{Tool call} & \texttt{query\_knowledge\_base(query="model substitutional doping in a crystal structure", file\_path=FDMNES\_manual.txt)} \\[4pt]
\textbf{Retrieved evidence} & The manual recommends the \texttt{Doping} keyword followed by the dopant atomic number and the index of the substituted site, for example \texttt{Doping 27 1}. It also states that the dopant is treated as a low-concentration substitution and that crystal symmetries are preserved. \\[4pt]
\textbf{Agent answer (excerpt)} & ``To model substitutional doping in a crystal structure using FDMNES, you can use the \texttt{Doping} keyword. This allows you to substitute a dopant species in place of another atom within the crystal structure.'' \\[6pt]

\midrule

\multicolumn{2}{l}{\textbf{Example 3}} \\[2pt]
\textbf{User query} & What are the default values for the energy range and step? \\[4pt]
\textbf{Tool call} & \texttt{query\_knowledge\_base(query="default values for the energy range and step", file\_path=FDMNES\_manual.txt)} \\[4pt]
\textbf{Retrieved evidence} & The retrieved manual passage states that the default energy range is $-5$ to $60$ eV with a step size of $0.5$ eV. It also notes that the \texttt{Range} keyword can be used to modify these values or define variable steps. \\[4pt]
\textbf{Agent answer (excerpt)} & ``The default values for the energy range and step in the FDMNES software are as follows: Energy Range: $-5$ to $60$ eV. Step: $0.5$ eV. These values can be modified by specifying the \texttt{Range} keyword in the input file.'' \\

\bottomrule
\end{tabular}
\end{table*}

To further test whether retrieved documentation was propagated into an executable simulation input, we ran a documentation-grounded XANES calculation in which the agent was asked to use the FDMNES manual to set the \texttt{Range} keyword before launching calculations for TiO$_2$ and Fe$_2$O$_3$. The recorded Gemini 2.5 Flash trace shows that the agent first loaded \texttt{FDMNES\_manual.txt} using HuggingFace embeddings, queried the knowledge base for the default \texttt{Range} values, retrieved the manual passage specifying an energy range from $-5$ to $60$ eV with a $0.5$ eV step, and then passed this value to \texttt{run\_xanes\_single} as \texttt{energy\_range=[-5.0, 0.5, 60.0]}. This trace provides a direct link between retrieved documentation and the numerical parameter used in the execution call. Importantly, the default numerical values for the \texttt{Range} keyword were not included in the user query, system prompt, or tool schema; they appeared only in the retrieved manual passage and in the subsequent execution argument. This example is therefore used as a parameter-propagation trace; routine ChemGraph executions can instead use the built-in ChemGraph mesh when a normalization-ready energy grid is desired.

These retrieval demonstrations should be interpreted as proof-of-capability examples rather than as a comprehensive benchmark of documentation-question answering accuracy. The repeated-execution study reported below evaluates consistency for one documentation query and one retrieved parameter, but it does not measure retrieval recall, answer accuracy across a broad parameter set, or robustness to ambiguous software-manual passages. Such evaluations would be valuable if documentation-grounded parameter assistance becomes a larger part of autonomous spectroscopy workflows.


\begin{table}[!h]
\centering
\caption{Documentation-grounded propagation of the FDMNES \texttt{Range} setting into the XANES execution call. Full trace is available at \url{https://github.com/vitorgrizzi/ChemGraph_xanes/blob/main/notebooks/rag_range_fill.txt}.}
\label{tab:range-provenance}
\renewcommand{\arraystretch}{1.2}
\small
\begin{tabular}{p{0.27\textwidth} p{0.62\textwidth}}
\toprule
\textbf{Step} & \textbf{Recorded trace evidence} \\
\midrule
Document loading & \texttt{load\_document(file\_path=FDMNES\_manual.txt, embedding\_provider=huggingface)} \\
Retrieval query & \texttt{query\_knowledge\_base(query="default values of the FDMNES Range keyword")} \\
Retrieved manual evidence & Default energy range from $-5$ to $60$ eV with a $0.5$ eV step. \\
Execution argument & \texttt{run\_xanes\_single(..., energy\_range=[-5.0, 0.5, 60.0], ...)} \\
\bottomrule
\end{tabular}
\end{table}

\subsection{Structure-file-based execution}
We first evaluated the file-based execution mode of the framework using the natural-language query: \textit{``Compute the XANES of the POSCAR file located at \path{/lcrc/globalscratch/vferreiragrizzi/agents/poscar_run/POSCAR} for Z absorber = 29.''} The agent must recognize that the query refers to a user-supplied atomistic input, extract both the file path and the absorber atomic number from the text, and map these quantities onto the structured arguments required by the XANES execution tool. In contrast to the documentation-grounded retrieval cases discussed above, this example probes the agent's ability to interpret an execution-oriented request and route a local structure file into the simulation workflow. 

\begin{figure}
    \centering
    \includegraphics[scale=0.4]{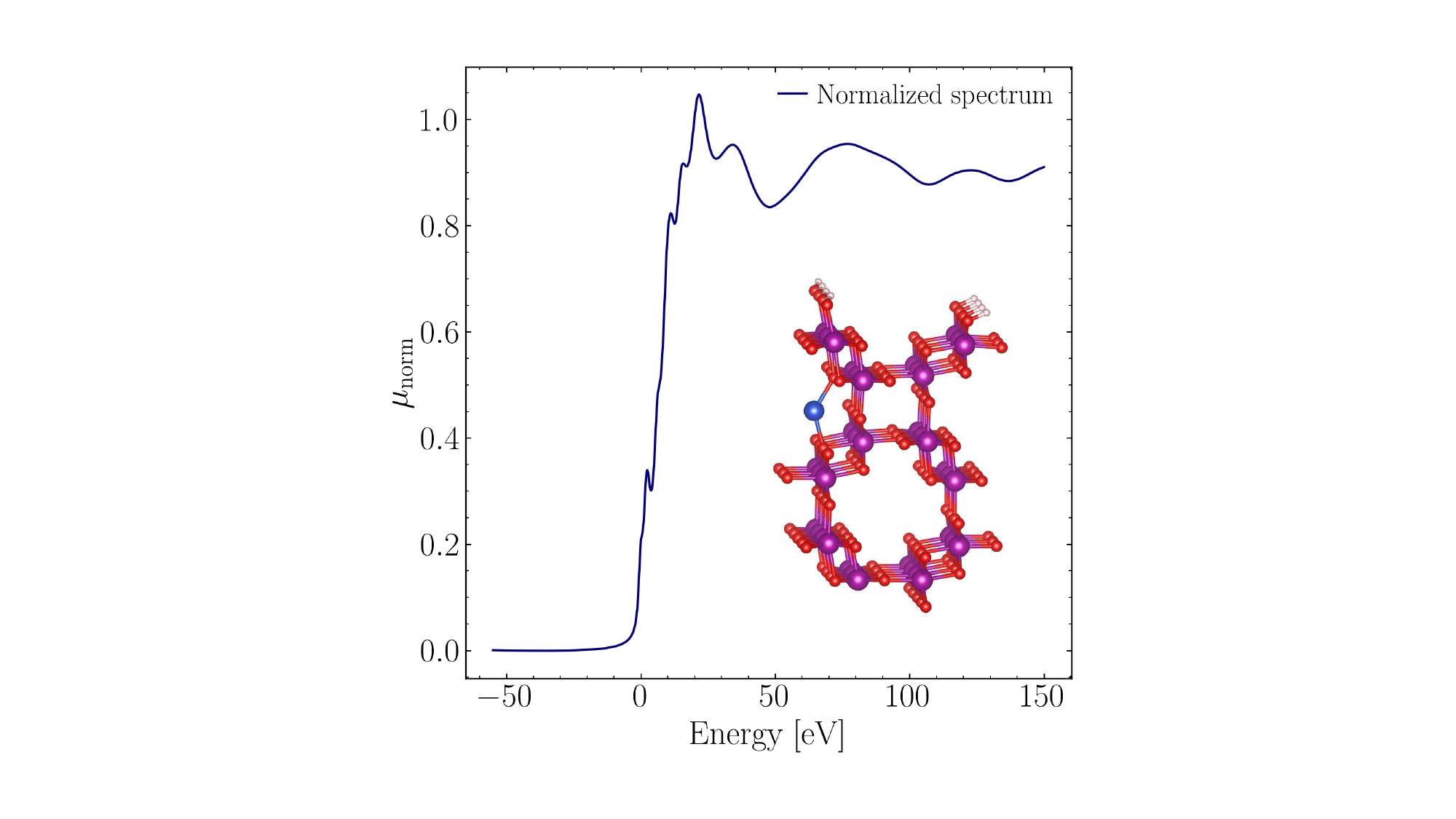}
    \caption{Normalized Cu K-edge XANES spectrum computed by setting the absorbing species to Cu on a MnO$_2$ (010) slab. The slab structure was constructed by cleaving MnO$_2$ along the (010) facet, and H atoms were added to passivate undercoordinated surface atoms. In the inset structure, the Cu species is shown in dark blue.}
    \label{fig:xanes_poscar}
\end{figure}

In this run, the POSCAR file was parsed into an \texttt{ase.Atoms} object, the corresponding FDMNES steering and input files were generated, the FDMNES calculation was launched, and the resulting convolution output was parsed, normalized, and plotted by the workflow. Figure~\ref{fig:xanes_poscar} shows the resulting normalized spectrum. The calculation was performed for the Cu K edge by setting the absorbing species to Cu ($Z=29$) on a MnO$_2$ (010) slab obtained by cleaving MnO$_2$ along the (010) facet and passivating undercoordinated surface sites with H atoms. The inset in Fig.~\ref{fig:xanes_poscar} shows the corresponding structure, with the Cu species highlighted in blue. No matching experimental spectrum or independent reference calculation was available for the specific MnO$_2$-based structure. This example is therefore presented as a demonstration of structure-file-based workflow execution, which does not require a composition-based search or database lookup, rather than as a validated spectral benchmark. This capability is important in practice because many computational spectroscopy workflows begin from local structures generated in prior DFT, molecular dynamics, slab-construction, or structural-relaxation steps rather than from canonical bulk entries in a materials database.

\subsection{Natural-language specification of a chemical system}
We next examined the natural-language interface using the query: \textit{``Compute the XANES for Ti in TiO$_2$.''} In contrast to the file-based example discussed above, this request does not provide an explicit local structure file. Instead, the agent must infer that the user is specifying a chemical system and absorbing species at a higher semantic level, identify the appropriate workflow for structure retrieval, and map the request onto the structured arguments needed for the downstream XANES tools. In practice, this means that the agent must interpret the chemistry-level request, issue a Materials Project query for TiO$_2$, identify Ti as the absorbing element, infer that this corresponds to absorbing-species atomic number $Z=22$, and pass these structured parameters to the downstream XANES workflow before triggering the same FDMNES input-generation and execution routines used in direct scripted mode. Spectral normalization and dataset expansion can then be applied through the separate post-processing helpers described above.

For the calculation reported here, the Materials Project query, using an energy-above-hull threshold of 0.001 eV/atom, resolved the formula-level request to anatase TiO$_2$ (Materials Project identifier \texttt{mp-390}). The identifier was retained in the calculation metadata to preserve the connection between the resulting spectrum and its source structure. This selection should not be interpreted as reproducing the experimental polymorph stability: rutile is the stable bulk phase under ambient conditions, whereas GGA and uncorrected r$^2$SCAN calculations can incorrectly place anatase below rutile in energy \cite{pohlmann2025stability}.

\begin{figure}
    \centering
    \includegraphics[scale=0.4]{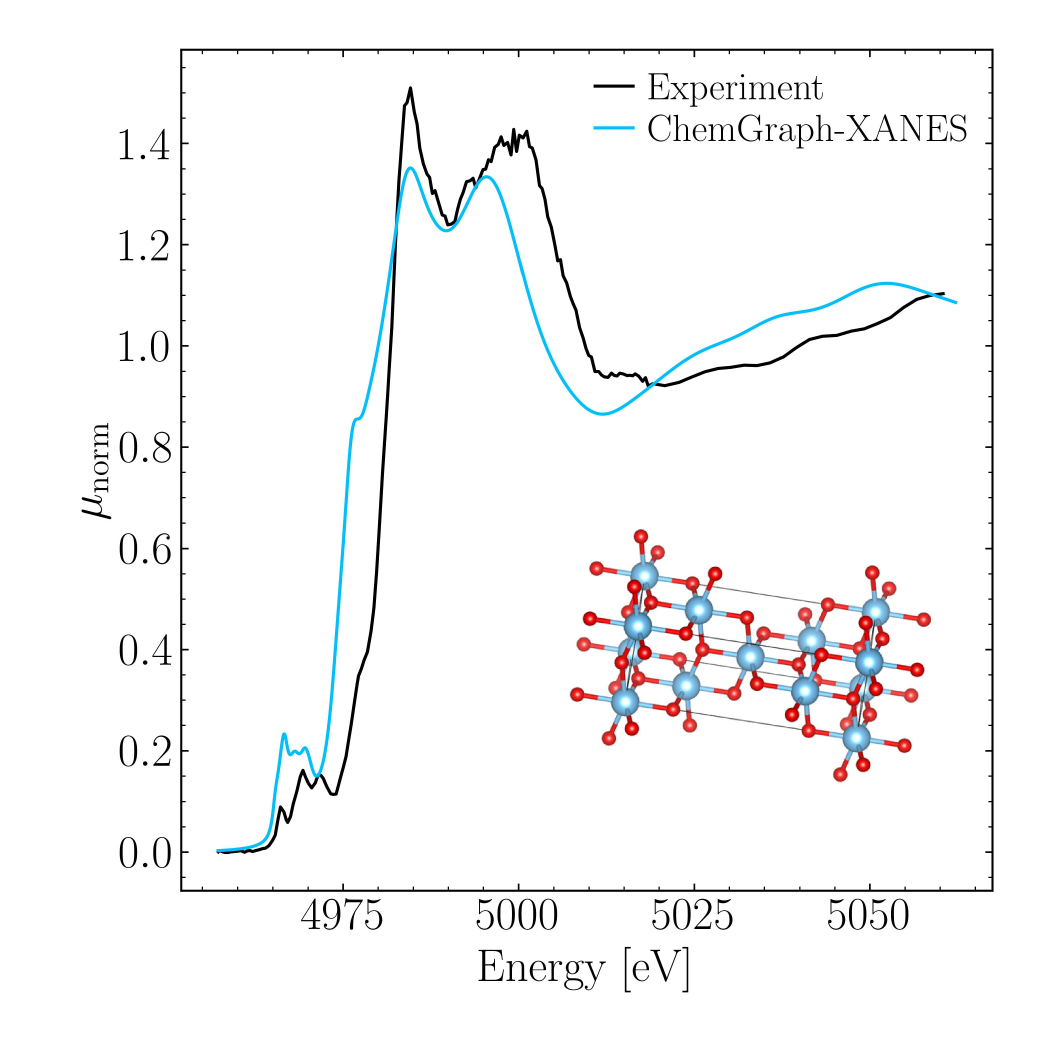}
    \caption{Calculated Ti K-edge XANES spectrum for anatase TiO$_2$ (\texttt{mp-390}) compared with the experimental anatase standard from the NIMS/SPring-8 XAFS database \cite{ishii}. Both spectra were normalized using the procedure described in the text. The calculation used a 6.0~$\mathrm{\AA}$ cluster radius in the FDMNES \texttt{Green} multiple-scattering mode with self-consistency and both dipole and quadrupole transitions enabled. A single rigid shift of $+4965.52$ eV was applied to align the highest-intensity main-edge peaks; no additional energy alignment or fitting was performed.}
    \label{fig:xanes_nl}
\end{figure}

Figure~\ref{fig:xanes_nl} shows the resulting normalized Ti K-edge spectrum for bulk TiO$_2$. The inset displays the corresponding crystal structure used in the calculation, with Ti atoms shown in light blue and O atoms in red. To assess the qualitative physical plausibility of the calculated spectrum, we compared it with an experimental Ti K-edge spectrum of anatase TiO$_2$ measured in transmission at SPring-8 BL14B2 and distributed through the NIMS Materials Data Repository \cite{ishii}. Because the FDMNES output used a relative energy scale, the calculated axis was shifted rigidly by $+4965.52$ eV to align the highest-intensity main-edge peaks; no other energy adjustment or spectral fitting was applied. The calculation used the \texttt{mp-390} structure, a 6.0~$\mathrm{\AA}$ cluster radius, self-consistency, and the \texttt{Green} multiple-scattering mode, with both dipole and quadrupole transitions included. The remaining differences may reflect the muffin-tin approximation associated with \texttt{Green}, finite-cluster effects, and non-optimized broadening \cite{joly2001x}. Full-potential calculations further demonstrate the sensitivity of the anatase pre-edge to $p$--$d$ hybridization and experimental geometry \cite{rossi2019x}.

Together, the file-based and composition-based examples show that different structure-specification modes can use the same scientific backend. Once the structure and calculation parameters have been resolved, both entry points use the same routines for FDMNES input generation and execution, with normalization available as a separate post-processing step. Structure selection from an underspecified composition-level request nevertheless remains part of the upstream orchestration, as examined in the repeated-execution results below.

\subsection{Repeated-execution consistency for representative workflows}

Under the repeated-execution protocol described in Methods, the composition-based and structure-file-based tasks each completed successfully in all ten trials. The RAG-grounded task completed both requested calculations in nine of ten trials. Across the 30 separate task executions, 29 completed end-to-end, corresponding to an observed aggregate success rate of 96.7\%. The task-specific outcomes are reported separately in Table~\ref{tab:success_rates} because the three workflows differ in complexity and number of requested calculations.

\begin{table}[ht]
\centering
\caption{Observed repeated-execution outcomes for three representative ChemGraph-XANES workflows under the fixed evaluation protocol described in Methods.}
\label{tab:success_rates}
\begin{tabular}{lcc}
\hline
Task & Completed trials & Observed success rate \\
\hline
Composition-based TiO$_2$ calculation & 10/10 & 100\% \\
POSCAR-based calculation & 10/10 & 100\% \\
RAG-grounded TiO$_2$/Fe$_2$O$_3$ calculation & 9/10 & 90\% \\
\hline
Overall & 29/30 & 96.7\% \\
\hline
\end{tabular}
\end{table}

In the unsuccessful RAG trial, the agent retrieved 14 TiO$_2$ structures instead of selecting the structure with the lowest energy above hull, as occurred in the other successful trials. It initialized a calculation for one structure, after which the trace recorded an error consistent with an attempt to process another structure in the same output directory. The Fe$_2$O$_3$ calculation completed successfully in that trial. Manual inspection showed that the default \texttt{Range} values retrieved from the manual were correctly propagated to the generated FDMNES input in all ten RAG trials, including the trial that failed during structure handling. Thus, parameter retrieval and propagation succeeded in 10/10 trials, whereas the complete two-material workflow succeeded in 9/10. The trace supports the interpretation that the failure occurred during multi-structure resolution and output management rather than during documentation retrieval or parameter propagation. This outcome also shows that schema validation constrains tool arguments but does not by itself determine how many structures an underspecified composition-level request should select. The scripts used for these experiments are available at \url{https://github.com/vitorgrizzi/ChemGraph_xanes/tree/main/examples/python_scripts}.

\subsection{Task-parallel TiO$_2$ polymorphs}
To evaluate the task-parallel execution and data-curation pathway, we performed a Ti K-edge XANES batch using the prompt template and calculation parameters described in Methods. The Materials Project retrieval tool returned 21 non-deprecated TiO$_2$ structures satisfying the specified energy-above-hull cutoff of 0.10~eV/atom. The returned structure list was used directly as the ensemble input, preserving the correspondence between the retrieved structures and submitted calculations. This produced 21 independent structure-level Parsl tasks, which were dispatched on the LCRC Improv cluster using ten one-node worker blocks. Each worker occupied one 128-core compute node and executed one FDMNES calculation at a time. Execution records confirmed that all ten worker nodes were active simultaneously and that the observed peak concurrency was ten FDMNES calculations. Tasks beyond the first ten remained in the Parsl queue and were assigned as workers became available.

\begin{table}[!h]
\centering
\caption{Configuration and outcome of the task-parallel TiO$_2$ XANES batch.}
\label{tab:tio2-parallel-batch}
\renewcommand{\arraystretch}{1.15}
\small
\begin{tabular}{lr}
\toprule
\textbf{Batch property} & \textbf{Value} \\
\midrule
Materials Project structures retrieved & 21 \\
Worker nodes used & 10 \\
Maximum concurrent calculations & 10 \\
Ensemble tasks executed & 21 \\
Successful calculations & 21 \\
Failed calculations & 0 \\
Existing outputs reused & 0 \\
Convolution outputs generated & 21 \\
\bottomrule
\end{tabular}
\end{table}

All 21 calculations completed successfully, and each run produced one convolution spectrum. The framework retained the retrieved structures in two forms: 21 individual CIF files named according to their Materials Project identifiers and compositions, and an \texttt{atoms\_db.pkl} file containing the complete collection as a pickle-serialized list of \texttt{ase.Atoms} objects, each of which retains its Materials Project identifier in its associated metadata. For each calculation, the framework retained the source structure, generated FDMNES input, execution logs, metadata, raw outputs, and convolution spectrum. These records preserve the link between each Materials Project identifier and its calculation parameters, execution status, and spectrum. A separate execution trace records the original request and the sequence of structure retrieval and ensemble execution. Manual inspection of the generated inputs for all 21 calculations confirmed that the absorber, edge, cluster radius, energy mesh, and FDMNES options were propagated as specified. Together, these records form a structure-linked XANES collection in which every spectrum can be traced to its source structure, calculation settings, and execution record. The complete collection is available at \url{https://github.com/vitorgrizzi/ChemGraph_xanes/tree/main/examples/parallel_task}.

\section{Discussion}
Automated XAS generation and spectroscopy workflow management predate the present work. The Materials Project/atomate effort established high-throughput FEFF-based XANES generation and structure-linked reference databases \cite{mathew2018high, zhang2018strategy}; Lightshow provides a common input-generation abstraction across several XAS codes \cite{carbone2023lightshow}; Corvus coordinates dependency-driven workflows across spectroscopy and electronic-structure packages \cite{story2019corvus}; and AiiDA QE Xspec provides provenance-aware workchains for core-level spectroscopy \cite{wang2026making}. ChemCrow established the broader utility of LLM agents equipped with chemistry tools \cite{m2024augmenting}, while the original ChemGraph framework supplies the general agent and tool infrastructure extended here \cite{pham2026chemgraph}. We therefore do not claim novelty for automated XAS, high-throughput execution, provenance tracking, or chemistry-tool-using agents individually. The contribution of ChemGraph-XANES is its integration in an FDMNES-specific interface in which documentation retrieval can supply numerical parameters to typed, schema-validated execution tools, while retaining an explicit link among retrieved evidence, final parameters, structure identity, execution records, and spectral outputs.

The main contribution of ChemGraph-XANES lies in formalizing a machine-actionable structure-to-spectrum workflow around established simulation software. Structure acquisition, absorber selection, FDMNES input generation, execution, parsing, normalization, and provenance tracking are organized within a common Python framework and exposed through typed tool interfaces. The LLM interprets requests and supplies schema-validated arguments, while FDMNES and deterministic backend routines perform the scientific calculations. Once structures and parameters have been resolved, local-file and chemistry-level requests therefore share the same execution pathway. Upstream interpretation is not fully deterministic, however: composition-level requests can leave structure cardinality unspecified, as demonstrated by the failed repeated RAG execution. The framework should consequently be understood as controlled orchestration rather than unrestricted or fully autonomous scientific decision-making.

General-purpose coding agents constitute another practical route to FDMNES automation. Given the FDMNES manual, shell access, and a correctly configured software environment, such an agent could likely construct scripts and complete the individual demonstrations presented here. ChemGraph-XANES therefore does not claim exclusive capability or benchmarked superiority. Its distinction lies instead in the execution contract: agent actions are restricted to predefined, schema-validated tools, scientific operations are delegated to a deterministic backend, and structures, resolved parameters, generated inputs, execution records, and spectra are retained through a common workflow. The constrained action space also reduces the orchestration burden because the model selects tools and supplies typed arguments rather than generating and debugging complete execution scripts. This design may reduce token usage and opportunities for unsupported actions, although token efficiency was not quantified here. Comparable safeguards can be built around general-purpose coding agents, but ChemGraph-XANES incorporates them directly into the XANES execution pathway. A systematic comparison remains an important direction for future work. 

The repeated RAG trials distinguish the documentation-grounded parameter-selection step from the broader orchestration workflow. The retrieved FDMNES \texttt{Range} setting was propagated correctly in all ten RAG trials, whereas the complete two-material workflow succeeded in nine; the single failure occurred later during multi-structure handling. Thus, the observed failure did not arise from documentation retrieval or parameter transfer. The significance of this example is not merely that the agent recovered a default numerical value from the manual. Rather, it shows that a simulation-relevant parameter can be made externally attributable to a retrieved documentation source and then propagated through a typed execution interface into a concrete XANES calculation. This distinction matters because one of the main risks in language-model-mediated scientific workflows is that parameter choices may otherwise appear plausible without being traceable to either the user, the software documentation, or an explicit default in the implementation. In ChemGraph-XANES, the retrieval trace therefore functions as a lightweight form of parameter provenance: the numerical value used in execution can be linked to a documented source rather than to unsupported free-form model generation.

The 21-structure TiO$_2$ batch extends the evidence beyond architectural compatibility with distributed execution. A single ensemble call created 21 structure-level tasks, all 21 calculations produced convolution spectra, and execution records showed a peak of ten concurrent FDMNES calculations. The retained structures, generated inputs, logs, metadata, raw outputs, and spectra form a traceable collection in which each spectrum remains linked to its source structure and calculation settings. This is directly relevant to comparative spectroscopy and ML-oriented dataset construction, for which hidden differences in structures, absorbers, simulation settings, or normalization can compromise later reuse. The experiment demonstrates task-parallel execution and structure-linked curation under one configuration, but it does not establish scaling efficiency because no serial reference or worker-count sweep was performed.

The present results retain several important limitations. First, this is a workflow study rather than a comprehensive benchmark of spectral fidelity; the examples do not systematically validate spectral accuracy across a broad materials set. Second, the repeated-execution evaluation comprises three fixed tasks, ten executions per task, one model, and one configuration. The resulting 29/30 outcome is therefore a task-specific observation rather than an estimate of general workflow reliability. Third, although the manual-derived \texttt{Range} setting was propagated in all ten RAG trials, this evaluates only one documentation query and one parameter; retrieval recall, answer accuracy across a broader parameter set, and robustness to ambiguous or incomplete manual passages were not measured. Finally, the TiO$_2$ collection was generated in one Parsl configuration. The observed concurrency demonstrates simultaneous execution, but no serial comparison, strong- or weak-scaling analysis, worker-count sweep, or systematic throughput study was undertaken.

There are also deliberate scope limitations in the current agent interface. The framework exposes a selected subset of FDMNES-relevant controls through validated schemas, rather than allowing the agent to modify arbitrary input keywords. This restriction limits flexibility for some advanced expert use cases, but it is central to the design: constraining the action space reduces the risk that an LLM will hallucinate plausible-looking but unsupported parameters, undocumented settings, or otherwise unintended simulation inputs. As LLMs and documentation-grounded tool-use methods improve, the set of exposed controls can be expanded progressively. In our view, however, such expansion should remain grounded in documentation and mediated through structured validation rather than unconstrained text generation. The current design therefore prioritizes trustworthy orchestration over maximal agent freedom.

Finally, although the present implementation centers on FDMNES, the architectural ideas are broader than a single spectroscopy backend. Structure handling, schema-validated orchestration, task organization, output parsing, and provenance-aware structure–spectrum linkage are conceptually separable from the specific simulation engine used to generate the spectra. This suggests a path toward future extensions in which similar adapters could be built for other XANES or spectroscopy codes while preserving a common workflow and data-curation layer. Such extensibility would be valuable for comparative-method studies and for broader efforts to build interoperable computational spectroscopy infrastructure.

\section{Conclusion}
We introduced ChemGraph-XANES, a documentation-grounded and schema-constrained framework for XANES simulation and data curation. The framework couples natural-language task specification with structured tool interfaces, deterministic FDMNES input generation, explicit post-processing helpers, and provenance-aware data organization. By acting as an orchestration layer over established physics-based software, ChemGraph-XANES preserves transparent execution logic while making common XANES workflows easier to access, reproduce, and reuse. Across ten separate executions of each of three representative workflows, 29 of 30 tasks completed end-to-end, and the manual-derived FDMNES \texttt{Range} setting was propagated correctly in all ten RAG trials. In the task-parallel evaluation, all 21 retrieved TiO$_2$ structures produced convolution spectra while retaining links to their structural inputs, execution parameters, logs, and raw outputs. These experiments demonstrate task-specific execution consistency and task-parallel collection generation, while broader reliability, retrieval, spectral-accuracy, and scaling behavior remain to be quantified. ChemGraph-XANES therefore provides a foundation for future high-throughput XANES campaigns, comparative spectroscopy studies, and ML-oriented structure-spectrum dataset generation, with an emphasis on constrained orchestration, reproducibility, and traceable data generation.

\section*{Conflicts of interest}
There are no conflicts to declare.

\section*{Data Availability}
The XANES-specific implementation used for the experiments in this work is available at \url{https://github.com/vitorgrizzi/ChemGraph_xanes}. This repository contains the research fork associated with the present manuscript. Ongoing development of the general ChemGraph framework is maintained in the upstream repository at \url{https://github.com/argonne-lcf/ChemGraph}. The qualitative documentation-retrieval examples are provided in \url{https://github.com/vitorgrizzi/ChemGraph_xanes/blob/main/notebooks/Demo_rag_agent_Argo.ipynb}, and the range-propagation trace is provided in \url{https://github.com/vitorgrizzi/ChemGraph_xanes/blob/main/notebooks/rag_range_fill.txt}. Scripts used for the repeated-execution experiments are available at \url{https://github.com/vitorgrizzi/ChemGraph_xanes/tree/main/examples/python_scripts}. The task-parallel prompt trace, Materials Project-derived structures, \texttt{xanes\_results.jsonl} summary, generated FDMNES inputs and logs, raw outputs, and convolution spectra are available at \url{https://github.com/vitorgrizzi/ChemGraph_xanes/tree/main/examples/parallel_task}.

\section*{Acknowledgments}
This work was supported by the U.S. Department of Energy (DOE), Office of Science (SC), Office of Basic Energy Sciences (BES), Division of Chemical Sciences, Geosciences, and Biosciences (CSGB), Catalysis Science Program at Argonne National Laboratory under contract no. DE-AC02-06CH11357. L.P. acknowledges U.S. DOE Office of Science Graduate Student Research (SCGSR) award under contract number DE-SC0014664. The authors gratefully acknowledge the computing resources provided on Swing and Improv, a high-performance computing cluster operated by the Laboratory Computing Resource Center (LCRC) at Argonne National Laboratory. This research used resources of the Argonne Leadership Computing Facility, which is a U.S. Department of Energy Office of Science User Facility operated under contract DE-AC02-06CH11357.

\bibliographystyle{unsrt}
\bibliography{ref}
\end{document}